\documentclass[]{article}
\pdfoutput=1
\usepackage{amssymb}
\usepackage[T1]{fontenc}
\usepackage{lmodern}
\usepackage{microtype}
\usepackage{graphicx}
\usepackage{caption}
\usepackage{subcaption}
\usepackage{booktabs}
\usepackage[nounderscore]{syntax}
\usepackage{microtype}
\usepackage{doi}
\usepackage{arxiv}

\usepackage[english]{babel}
\usepackage[utf8]{inputenc}
\usepackage{url}

\DeclareMathSymbol{\Alpha}{\mathalpha}{operators}{"41}
\DeclareMathSymbol{\Beta}{\mathalpha}{operators}{"42}
\DeclareMathSymbol{\Epsilon}{\mathalpha}{operators}{"45}
\DeclareMathSymbol{\Rho}{\mathalpha}{operators}{"50}

\def\uschema{U-Schema}
\def\skiql{SkiQL}

\usepackage{color}

\usepackage{graphicx}
\usepackage{xcolor}
\usepackage{listings}

\definecolor{javapurple}{rgb}{0.5,0,0.35}
\definecolor{delim}{RGB}{20,105,176}
\colorlet{punct}{red!60!black}
\colorlet{numb}{magenta!60!black}

\lstdefinelanguage{json}{
    basicstyle=\scriptsize\ttfamily,
    showstringspaces=false,
    breakatwhitespace=true,
    breaklines=true,
    morekeywords={_id,_type,count,schema},
    literate=
     *{0}{{{\color{numb}0}}}{1}
      {1}{{{\color{numb}1}}}{1}
      {2}{{{\color{numb}2}}}{1}
      {3}{{{\color{numb}3}}}{1}
      {4}{{{\color{numb}4}}}{1}
      {5}{{{\color{numb}5}}}{1}
      {6}{{{\color{numb}6}}}{1}
      {7}{{{\color{numb}7}}}{1}
      {8}{{{\color{numb}8}}}{1}
      {9}{{{\color{numb}9}}}{1}
      {:}{{{\color{punct}{:}}}}{1}
      {,}{{{\color{punct}{,}}}}{1}
      {\{}{{{\color{delim}{\{}}}}{1}
      {\}}{{{\color{delim}{\}}}}}{1}
      {[}{{{\color{delim}{[}}}}{1}
      {]}{{{\color{delim}{]}}}}{1}
}

\lstdefinelanguage{cypher}{
        morecomment=[l][\color{gray}]{//},
        morestring=[b][\color{blue}]\",
        morestring=[b][\color{blue}]\',
        morekeywords=[2]{WITH,WITH,AS,MATCH,WHERE,IN,ALL,RETURN,RETURN}
    sensitive=false,
        keywordstyle=[2]\color{javapurple}
}

\lstdefinelanguage{skiql}{
        morecomment=[l][\color{gray}]{//},
        morestring=[b][\color{olive}]\",
        morestring=[b][\color{olive}]\',
        morekeywords=[2]{ENTITY,PROPERTIES,ALL,FROM,TO, REF, REL, ANY,
    AGGR,ATTRIBUTES,REFERENCES,AGGREGATIONS,NON,SHARED,SPECIFIC, COUNT,
    HISTORY, ONLY, BEFORE, AFTER, BETWEEN},
    sensitive=false,
        keywordstyle=[2]\color{javapurple}
}

\title{\skiql{}: A Unified Schema Query Language\thanks{This work
 has been funded by the Spanish Ministry of Science and Innovation, project grant
 PID2020-117391GB-I00.}~$^,$\thanks{Formatted for arXiv.org.}}

\author{  \href{https://orcid.org/0000-0002-3835-9428}{\includegraphics[scale=0.06]{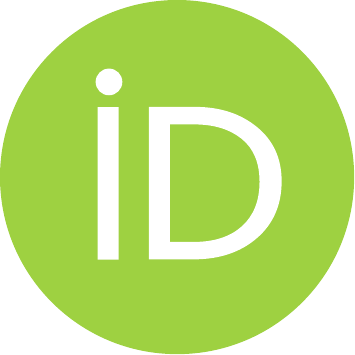}\hspace{1mm}Carlos J. Fernández Candel} \\
  Faculty of Computer Science\\
  University of Murcia\\
  Murcia, Spain\\
        \texttt{cjferna@um.es} \\
        \And \href{https://orcid.org/0000-0003-4685-6659}{\includegraphics[scale=0.06]{orcid.pdf}\hspace{1mm}Jesús J. García-Molina} \\
  Faculty of Computer Science\\
  University of Murcia\\
  Murcia, Spain\\
        \texttt{jmolina@um.es}\\
        \And \href{https://orcid.org/0000-0001-9313-008X}{\includegraphics[scale=0.06]{orcid.pdf}\hspace{1mm}Diego Sevilla Ruiz} \\
  Faculty of Computer Science\\
  University of Murcia\\
  Murcia, Spain\\
        \texttt{dsevilla@um.es}\\

}

\renewcommand{\shorttitle}{\skiql{}: A Unified Schema Query Language}

\begin{document}
\maketitle

\begin{abstract}

  Most NoSQL systems are schema-on-read: data can be stored without first
  having to declare a schema that imposes a structure. This schemaless
  feature offers flexibility to evolve data-intensive applications when
  data frequently change.
  However, freeing from declaring schemas does not mean their absence, but
  rather that they are implicit in data and code. Therefore, diagramming
  tools similar to those available for relational systems are also needed
  to help developers and administrators designing and understanding NoSQL
  schemas.

  Visualizing diagrams is not practical if schemas contain hundreds of
  database entities, and exploration or query facilities are then needed.
  In schemaless NoSQL stores, data of the same entity can be stored with
  different structure (e.g.,~non-uniform types and optional fields), which
  can increase the difficulty of having readable diagrams.

  NoSQL schema management tools should therefore have three main
  components: schema extraction, schema visualization, and schema query.
  Since that there exists four main NoSQL data models,
  it is convenient that such tools can be built on a generic data model so
  that they provide platform-independence (of data models and data stores)
  to query and visualize schemas. With the aim of favoring the creation of
  generic database tools, the authors of this paper defined the \uschema{}
  unified data model that integrates the four main NoSQL data models as
  well as the relational model.

  This paper is focused on querying NoSQL and relational schemas which are
  represented as \uschema{} models. We present the \skiql{} language
  designed on \uschema{} to achieve a platform-independent schema query
  service. \skiql{} provides two constructs: schema-query and
  relationship-query. The former allows to obtain information of entity or
  relationship types, and the latter that of the aggregations or references
  (relations among types). We will show how \skiql{} was evaluated by
  calculating well-known metrics for languages as well as using a survey
  with developers with experience in NoSQL.
\end{abstract}


\section{Introduction}
\label{sec:introduction}

When data structure frequently changes, as occurs in modern applications,
the need of formally specifying database schemas hampers agile development.
Thus, most NoSQL systems are schemaless and developers are not forced to
declare schemas prior to store data. However, not having to declare schemas
does not mean their absence, but that they are implicit in stored data or
application code. Managing data always requires to design, create and
evolve database schemas by developers or administrators. Developers must
always keep in mind the schema to write code, and administrators have to
know the schema in order to perform common tasks as optimizing queries.
Therefore, database tools to manage schemas are as essential for NoSQL
systems as they have been for relational systems.

Data modeling tools offer diagramming facilities to help developers to
create and understand schemas. In the case of schemaless NoSQL stores, they
should support the extraction of schemas as they are not declared, as well
as addressing the variety of data models that are currently used. Moreover,
the need for schema query languages is even greater than in relational
databases, as discussed below.

Diagramming is not practical when schemas include many entities
(e.g.,~tables in a relational schema), and a schema query language is then
convenient. In relational databases, this language can be the proper SQL
since that the standard SQL-92 specifies how information on schemas could
be represented in form of tables.
In the case of schemaless NoSQL stores, no data checking against the schema
is performed, and data of the same entity can be stored with different
structures, which we will refer to as \textit{structural variations}.
Variations can complicate the visualization of a readable schema, in
particular if most of entities have variations or some entity has a large
number of variations. In fact, thousands of variations for a database
entity have been found in datasets of some domains, such as DBpedia or
molecular biology, so that using query languages is proposed
in~\cite{wang-schema2015}. In~\cite{alberto-erforum2017}, a visual notation
was proposed for NoSQL schemas, and that work evidenced that diagramming is
not useful when information on variations is desired, although the number
of variations is small, and therefore a query language is essential to
inspect NoSQL schemas.

There are four main categories of NoSQL systems: columnar, document,
key-value, and graph. The lack of a data model standard for these four
categories causes that NoSQL systems of the same kind can significantly
vary in their features and in the structure of the data. This variety has
motivated that existing relational modeling tools (e.g.,~Erwing and
ERStudio) are evolving to support several models and new multi-model tools
are appearing (e.g.,~Hackolade). Another fact that supports the building of
multi-model systems is the emergence of \textit{polyglot persistence}: one
kind of database does not fit all the needs~\cite{stonebraker-blog-2015}.

With the aim of favoring the creation of generic database tools and
languages, the authors of this paper defined the \uschema{} unified data
model that integrates the four main NoSQL data models and the relational
model, as described in~\cite{metamodel2021}. In that paper, we implemented
\uschema{} schema extractors for NoSQL and relational stores. In this
paper, we will focus on logical schema querying. We will present the
\skiql{} schema query language defined on the \uschema{} unified data
model. Using \uschema{}, \skiql{} will be a language that is independent of
a particular data model and data store.

\skiql{} offers two constructs that are applicable to any database schema
represented as a \uschema{} model: \emph{entity-query} and
\emph{from\_to-query}. The former allows to obtain information from an
entity type, and the latter returns a sub-schema that includes the
relationships (aggregations and references) between entity types specified
in the query. Additionally, \skiql{} includes the \emph{relationship-query}
construct aimed to obtain information from a relationship type, and it is
only applicable in the case of \uschema{} schemas coming from graph stores.
The \skiql{} engine executes queries on the schema and builds a new schema
with the selected elements. This result schema is returned in form of a
diagram. It should be noted that the work is focused on the query language
rather than the visualization of the query results. \skiql{} is independent
of the visual representation of the result returned by queries. An
evaluation of the language has been performed through an experiment with
experienced NoSQL researchers and developers, and calculating some
well-known metrics for domain-specific languages.

\paragraph*{Research contributions} The main contributions of this work are
the following.

\begin{itemize}
\item As far as we know, \skiql{} is the first proposal of a generic
  language to query schemas. \skiql{} is applicable to relational and NoSQL
  logical schemas thanks to the usage of \uschema{} as the pivotal data
  model. Relationship types can be queried in graph stores.

\item The creation of a schema query language for NoSQL systems has only
  been addressed for document stores in~\cite{wang-schema2015}. In that
  work, the interest of querying large document schemas is illustrated with
  two query examples expressed in a SQL-like language, but the design and
  implementation of a complete language is not considered. \skiql{} is a
  more complete language that allows to perform a greater set of queries
  over the schema.

\item \skiql{} language takes advantage of the \uschema{} characteristics
  to allow entity variations to be explored and the schemas to be traversed
  through aggregation and reference relationships between entity types.
  Also, it support a more traditional approach using union types collapsing
  variation information.

\end{itemize}

This article has been organized as follows. Related work is discussed in
the following section. Next, the diagramming notation designed to represent
query results is presented by using a running example schema. Then,
\skiql{} syntax and semantics is describe in detail. Finally, the work
performed to evaluate the \skiql{} features is presented, and some
conclusions and further works are exposed.

\section{Related Work}
\label{sec:relatedwork}

A \emph{data dictionary} is managed in relational database systems to
register metadata on the stored data. This metadata includes the database
schema and information about physical implementation, security, and
programs (i.e.,~triggers), among other aspects. The SQL-92 standard
specifies the structure of data dictionaries in form of tables and views.
SQL can therefore be used to recover information from data dictionaries,
and queries can return information on the logical schema, e.g., tables
without primary keys, or tables that are not referenced by foreign keys. In
fact, data dictionary information is used to visualize relational schemas
in data modeling and metadata management tools. In~\cite{dataedo-web},
useful queries to explore schemas in several popular databases can be
found. In sections~\ref{sec:QT} and~\ref{sec:QR}, we will show how these
kind of SQL queries on data dictionaries can be expressed with \skiql{} in
a more concise and simple way.

In the case of NoSQL systems, a schema native query facility is only
provided in systems in which a schema may or must be declared, such as
OrientDB\footnote{OrientDB Community Webpage:
  \url{http://orientdb.com/orientdb}.} or Cassandra.\footnote{Cassandra
  Webpage: \url{http://cassandra.apache.org}.} OrientDB is a multi-database
system (graph and document) that allows to work schemaless or with a
declared schema. When developers declare the database schema, they can
issue SQL queries on the schema, indexes, or storage. These queries
returning information in form of tables~\cite{orientdb-metadata-web}.
Cassandra is a popular columnar store that offers a schema query support
similar to OrientDB. In both systems, queries are issued on a physical
schema.

To our knowledge, the ability to perform queries on entity variations is
only addressed in a work by Wang et al.~\cite{wang-schema2015}. These
authors observed that well-known datasets have entities with tens of
thousands of variations, which largely complicates the extraction and
visualization of schemas. Their work focused on document databases, in
particular MongoDB\footnote{MongoDB Webpage:
  \url{http://www.mongodb.com}.}, and a document schema management
framework is presented to tackle the problem. This framework includes a set
of utilities aimed to extract, persist, and query schemas. Extracted
schemas are recorded in a data structure defined as part of the work to
facilitate queries on schemas: \emph{eSiBuTree} trees. A SQL-like language
is proposed to express queries, but the authors only show a couple of
examples:~(i)~to check if a particular variation, which is specified by a
list of properties, exists for a given entity, and~(ii)~to find which
variations of a given entity have a concrete property. A limitation of this
proposal is that queries are issued only on entities because relationships
between entities are not inferred: references and entities for embedded
objects are not inferred. The query language suggested, as far we know, has
not been completely defined and implemented yet. Our proposal differs from
the approach of~\cite{wang-schema2015} in the following aspects:

\begin{itemize}
\item We present a complete query language, with its corresponding
  implementation.
\item It is multi-model supporting the four most popular NoSQL systems, as
  well as relational systems.
\item It has been devised keeping in mind the existence of relationships
  both between entities and between entity variations.
\item Queries can return information on entities, variations, or
  relationships.
\end{itemize}

Other schema query solutions are available, but they do not support
relationships or entity variations. Variety\footnote{Variety repository:
  \url{http://www.github.com/variety/variety}.} is a database tool
developed for MongoDB, which provides support to analyze schemas. The
queries are expressed in Javascript code, instead of using a language
tailored to query schemas. Apache Drill\footnote{Apache Drill Webpage:
  \url{http://drill.apache.org}.} is a SQL-based query engine that support
access to any kind of database (relational, NoSQL, Hadoop, etc.). Drill
represents the extracted schema in form of relational schema, so that SQL
queries can be issued on it.

Before creating \skiql{}, we experimented with the Cypher language to query
NoSQL schemas in~\cite{carlos-jisbd2019}. For this, we injected schemas
inferred from MongoDB databases into Neo4j\footnote{Neo4j Webpage:
  \url{http://www.neo4j.com}.} databases and used Cypher to write queries
on graph schemas. The resulting graphs were visualized with the Neo4j
Browser tool.
In our experiment, we observed that the queries were normally large and
difficult to write, which could even cause long execution times.
We estimated that the average number of Cypher LoC for simple queries was
about~8. This size is mainly due to the need of expressing the path to be
traversed, and storing the visited nodes in variables. It should be noted
that the size estimation was performed for simple queries. We then decided
to build a domain-specific language (DSL) tailored to query schemas
represented with the \uschema{} unified metamodel. In
Section~\ref{sec:metrics}, two examples of schema queries are expressed in
Cypher and in our DSL to evidence the greater length and complexity of the
Cypher queries (listings~\ref{lst:cypherQ1} and~\ref{lst:cypherQ4}).

\section{The \uschema{} Unified Data Model\label{sec:uschema}}

\skiql{} aims to query logical database schemas represented as instances of
the \uschema{} unified data model presented in~\cite{metamodel2021}. This
data model provides a uniform representation for schemas extracted from
relational databases and the four more popular NoSQL stores: columnar,
document, key-value, and graph. Building \skiql{} on top of \uschema{}, we
allowed us to have a query language independent of the data model. In this
section, we will describe the \uschema{} data model in sufficient detail to
understand the rest of the paper.

\begin{figure*}[!htb]\centering
  \includegraphics[width=\textwidth]{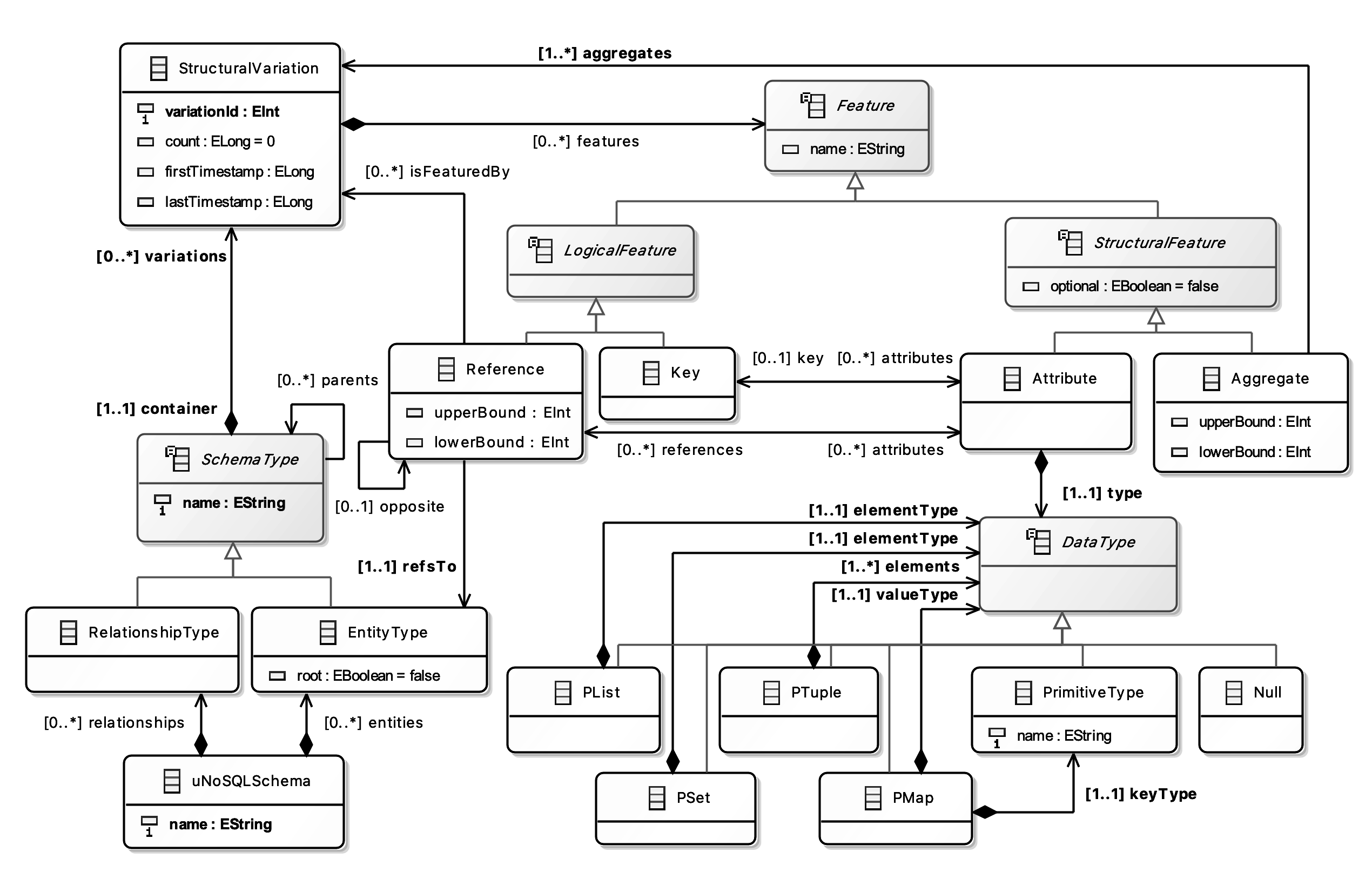}
  \caption{\uschema{} Data Model.\label{fig:uschemametamodel}}
\end{figure*}

\uschema{} has been defined as a metamodel whose instances (i.e., models)
are schemas. Figure~\ref{fig:uschemametamodel} shows the \uschema{}
metamodel. To understand the metamodel, it is convenient to consider that
the four categories of NoSQL stores are classified in two groups depending
on the kind of relationship that is prevalent to organize data. In
``aggregate-based data models,'' databases store semi-structured data which
form aggregation hierarchies, and references are established by using
object identifiers. Columnar, document, and key-value systems are based on
aggregation. Instead, graph-based systems are organized as a graph of
objects connected through arcs denoting a binary relationship between the
entity types to which connected objects belong, and aggregation is not
supported. In aggregate-based systems, objects stored are instances of
entity types, while both entity and relationship types can be instantiated
in graph-based systems.

A \uschema{} model (i.e., a schema) is formed by a set of entity types and
relationship types. \emph{Entity types} represent real-world entities whose
data are stored, and \emph{Relationship types} specify relationships
between entities. Relationship types only exist in schemas coming from
graph stores. We will use the ``schema type'' term to refer to both entity
type and relationship type.

Entity types can be root or aggregate. The root entity types are those
whose instances are not embedded into other objects, while the instances of
aggregate entity types are embedded into a root object or another embedded
object. Aggregate entity types allow hierarchies of aggregation proper of
semi-structured objects, and they are not supported for graph systems.

Types in \uschema{} have a name and a set of \emph{structural variations}.
These variations are characterized by a set of \emph{features} that can be
of two kinds: logical, and structural. Variations can be identified by a
numeric value which range from $1$ to the number total of variations.

\emph{Structural features} denote properties that hold the values of the
database objects, which can be \emph{attributes} or \emph{aggregates}.
\emph{Relationship types} have attributes but not aggregates.
\emph{Attribute}s and \emph{Aggregate}s are name-type pairs. For
attributes, the type can be either simple (\textit{Number},
\textit{String}, \textit{Boolean}) or structured (\textit{Set},
\textit{List}, \textit{Map}, and \textit{Tuple}). In the case of
aggregates, the type is a variation of an aggregate entity type, or a
collection of variations of the same aggregate entity type.

\emph{Logical features} denote properties that hold an object identifiers.
They are formed by one or more attributes and can be of two types: an
object \emph{key} or a \emph{reference} to another object. In the case of
graph schemas, each reference is featured by one or more variations of a
relationship type: they model the set of features of the reference.

Given an entity type $e$ with $n$ variations, its properties can be
classified into \emph{common} (a.k.a. \emph{shared} or \emph{required}) and
\emph{optional} (a.k.a. \emph{non-shared}) depending on whether the
property is or not present in all the variations of $e$. In turn, an
optional property can be \emph{specific} to only one variation or to $m$
variations of $e$, where $m<n$.

\section{Visualization of \skiql{} Query Results\label{sec:visualization}}

Before explaining the syntax and semantics of \skiql{}, we will present the
graphical notation devised to show the query results. The notation will be
described showing how various NoSQL schema types and very simple query
results are represented.

\subsection{Kinds of NoSQL Schemas\label{sec:kindschemas}}

Most NoSQL schema inference
approaches~\cite{wang-schema2015,klettke-schema2015} do not extract
variations or relationships between entities. Thus, the schema notion
considered is the set of \textit{union entity types}, where each type is
formed by the union of all the features that are present in its variations.
Instead, a \uschema{} schema contains a set of schema types, with their
variations, and the relationships (aggregation and references) between
variations or types.
Next, we define some sub-schemas and schema views that are of interest in
providing useful information on the stored data structure.

\paragraph{Type-Variations Sub-schema}
It contains a schema type and all its variations. All the properties of a
variation are considered pairs formed by its name and type, and the
cardinality is also included for aggregations and references.

\paragraph{Union Type}
It is a reduced view of a type-variations sub-schema that results of
gathering all the variations of the type into a single variation. The set
of features of such a variation is the union of the set of features of the
gathered variations. Obtaining the union of features requires making a
decision to resolve feature name collisions: the same feature name is
associated to different types in different variations of the same entity
type. We decided to use union data types, i.e.,~a feature can have multiple
types.

\paragraph{Simple Schema of Union Types} Joins all the union types of a
schema. As mentioned above, this kind of schema is the result commonly
obtained in the extraction approaches for document stores, such
as~\cite{klettke-schema2015} and~\cite{wang-schema2015}. Actually, it is
not a schema, but a reduced view of a NoSQL complete schema.

\paragraph{Complete Schema of Union Types} It also joins all the union
types, but relationships between entity types are also added. It is also a
view of the complete \uschema{} schema. This kind of schema corresponds to
the logical schema typically used for relational databases.

\subsection{Visualization of Complete Schemas\label{sec:diagrams}}

\begin{figure*}[!h]
\centering
\begin{minipage}[t]{.45\textwidth}
\begin{lstlisting}[
        frame=tb,
        basicstyle=\fontsize{8}{9}\selectfont\ttfamily,
        captionpos=b,
        numbers=left,
        language=JSON
]
// User Collection
{
  _id: 178,
  name: "Brian",
  surname: "Caldwell",
  email: "brian_caldwell@gmail.com",
  address: {
    city: "Aylesbury",
    street: "Fairfax Cres",
    number: 6,
    postCode: 30760
  },
  watchedMovies: [
    {
      stars: 4,
      movie_id: 202
    }
  ],
  favoriteMovies: [
    202,
    267,
    378
  ]
}
\end{lstlisting}

\begin{lstlisting}[
        frame=tb,
        basicstyle=\fontsize{8}{9}\selectfont\ttfamily,
        captionpos=b,
        numbers=left,
        language=JSON
]
// Movie Collection
{
  _id: 202,
  name: "The Matrix",
  year: 1999,
  genre: "Science Fiction"
}
\end{lstlisting}

\end{minipage}
\qquad
\begin{minipage}[t]{.45\textwidth}

\begin{lstlisting}[
        frame=tb,
        basicstyle=\fontsize{8}{9}\selectfont\ttfamily,
        captionpos=b,
        numbers=left,
        firstnumber=25,
        language=JSON
]
// User Collection (cont'd)
{
  _id: 156,
  name: "Allison",
  email: "allison@gmail.com",
  address: {
    city: "Aylesbury",
    street: "Lott Meadow",
    number: 8
  },
  "watchedMovies": [{
      stars: 3,
      movie_id: 202
    }, {
      stars: 5,
      movie_id: 295
  }]
}
\end{lstlisting}
  \includegraphics[width=\textwidth]{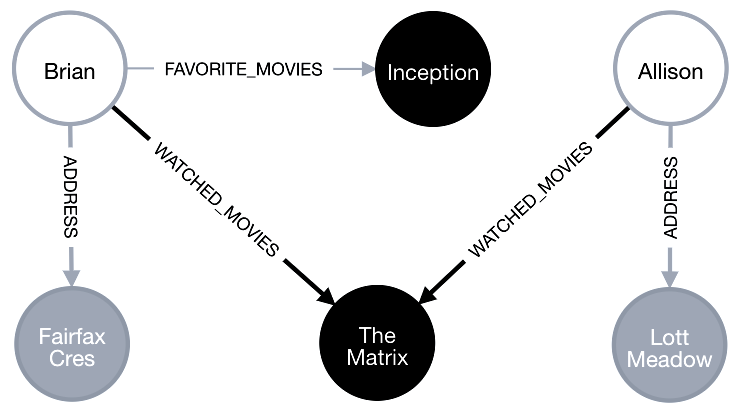}
\end{minipage}%
\caption{Running example for aggregate and graph stores.}
\label{fig:examples}
\end{figure*}

A simple \textit{User Profile} database will be used as a running example
throughout this paper. The database records data on users subscribed to a
streaming service: personal data, watched movies and favorite movies;
addresses will be separated from the rest of personal data. We will suppose
that this database will be stored both in an aggregate-based system
(e.g.,~MongoDB or Cassandra) and a graph system (e.g.,~Neo4j), and both
stores will be called ``UP-aggregate'' and ``UP-graph.''
Figure~\ref{fig:examples} shows the running example for the two stores.

In Figure~\ref{fig:examples}, two \textit{User} objects and one
\textit{Movie} object of ``UP-aggregate'' are shown in form of JSON
documents stored in MongoDB. A \textit{User} aggregates an \textit{Address}
object and an array of \textit{WatchedMovies} objects, and also holds an
array of references to \textit{Movie} objects that records the user's
favorite movies. Each \textit{WatchedMovies} holds a reference to the
\textit{Movie} watched by the user and the number of stars of the user's
score. In the case of ``UP-graph,'' \textit{Address}es and
\textit{WatchedMovies} are also connected to \textit{User}s through
reference relationships, as shown in Figure~\ref{fig:examples}.

\textit{User} and \textit{Address} have two variations, while
\textit{Movie} and \textit{WatchedMovies} have only one, i.e.,~there is not
structural variability for these two entities. \textit{User} and
\textit{Address} variations will be commented below when explaining the
schema diagrams.

Table~\ref{table:uschema-viewer-mappings} shows the mapping between
\uschema{} and graphical notation elements. Query results are visualized as
diagrams in which there are two kinds of nodes:

\begin{itemize}
\item Schema types are represented as boxes with two compartments: «entity
  type» or «relationship type» stereotypes appear on the upper one, and
  the type name on the lower one; different colors are used to make it
  easier to identify the types of nodes: light yellow for root entity
  types, light gray for aggregate entity types, and light blue for
  relationship types.

\item Variations are represented as white boxes with two compartments: the
  variation name and identifier appear in the upper one, and the list of
  features in the lower one.
\end{itemize}

These nodes are connected by means of four kinds of arrows as indicated in
Table~\ref{table:uschema-viewer-mappings}:~(i)~schema type to
variation,~(ii)~variation to aggregated variation,~(iii)~variation to
referenced entity, and~(iv)~reference to the relationship type that
specifies it.

Features are prefixed with ``+'', ``?'', and ``-'' symbols to indicate if
they are shared, non-shared, or specific. In the case of aggregation and
reference arrows, this prefix is followed for the cardinality specification
before the property name: ``[0..1]'' (zero to one), ``[1..1]'' (only one),
``[0..*]'' (zero to many), and ``[1..*]'' (one to many). It is worth noting
that references and aggregations that belong to variations present in the
query result but are not part of the set of relationships returned, will be
shown in the lower compartment of its variation. They will appear in the
same way as features, but indicating the kind of relationship (``-{}-'' or
``<>-'') and the cardinality.

\begin{table*}[]
\begin{center}
\begin{tabular}{lp{.6\textwidth}}
\toprule
\textbf{\uschema{} Elements} & \textbf{Graphical Notation Elements} \\
\midrule

\textit{Entity Type} & Box stereotyped with «Entity Type»\\[.9ex]

\textit{Relationship Type} & Box stereotyped with «Relationship Type»\\[.9ex]

\textit{Structural Variation} & Box with two compartments, one to specify
                                the variation and another to include its
                                list of feature specifications\\[.9ex]

\textit{Variation belongs to a schema type} & Dashed arrow from an Entity
                                              box to a variation box\\[.9ex]

\textit{Attribute} & Name and type separated by colon inside a variation
                     box\\[.9ex]

\textit{Key} & Prefix ``Key'' followed by the key's name and type inside a
               Variation box\\[.9ex]

\textit{Reference} & Blue arrow directed from referencing entity to
                     referenced entity, and labeled with the cardinality
                     and reference name\\[.9ex]

\textit{Aggregation} & Red arrow directed from aggregate entity to
                         aggregated entity, and labeled with the
                         cardinality and aggregation name, and
                         decorated with a filled diamond at the
                         aggregate class end\\[.9ex]

\textit{Variation featuring a reference} & Dashed arrow from the middle of
                                           a reference arrow to a variation
                                           box of a relationship type\\[.9ex]

\bottomrule
\end{tabular}
\end{center}
\caption{Mapping between \uschema{} and graphical
  notation.\label{table:uschema-viewer-mappings}}
\end{table*}

\begin{figure*}[!htb]
  \centering
  \includegraphics[width=\textwidth]{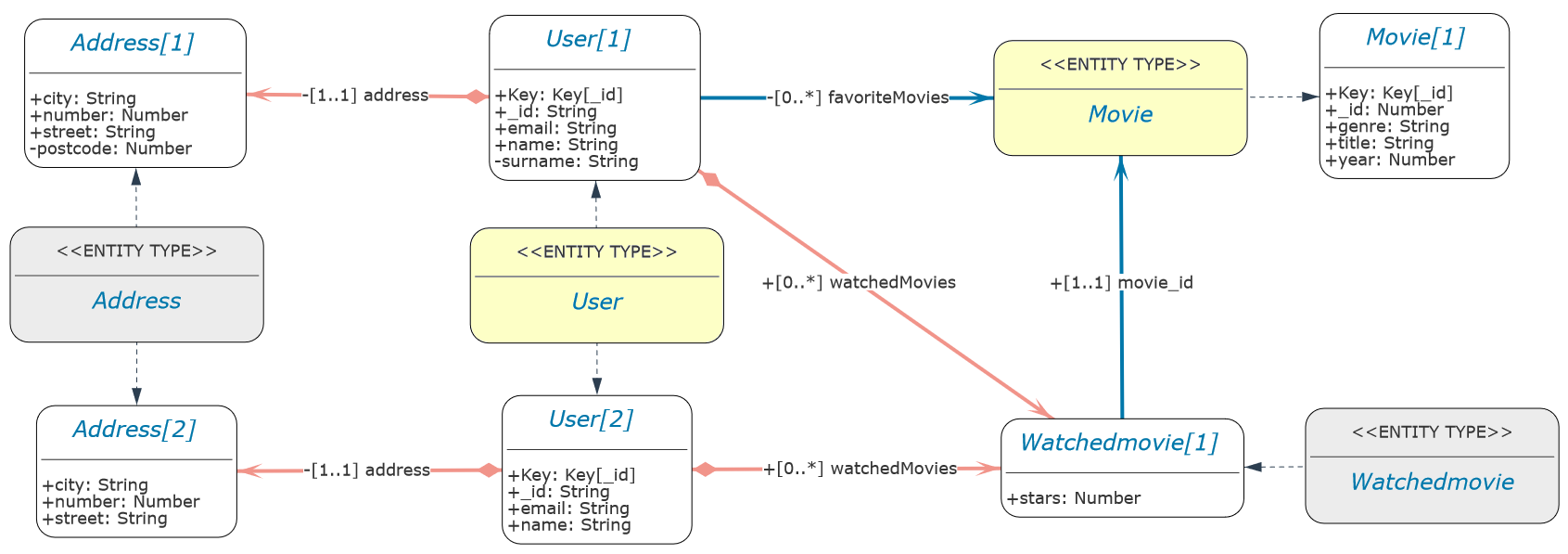}
  \caption{\uschema{} complete schema for ``UP-aggregate.''}
  \label{fig:q1DocAll}
\end{figure*}

\begin{figure*}[!htb]
  \centering
  \includegraphics[width=\textwidth]{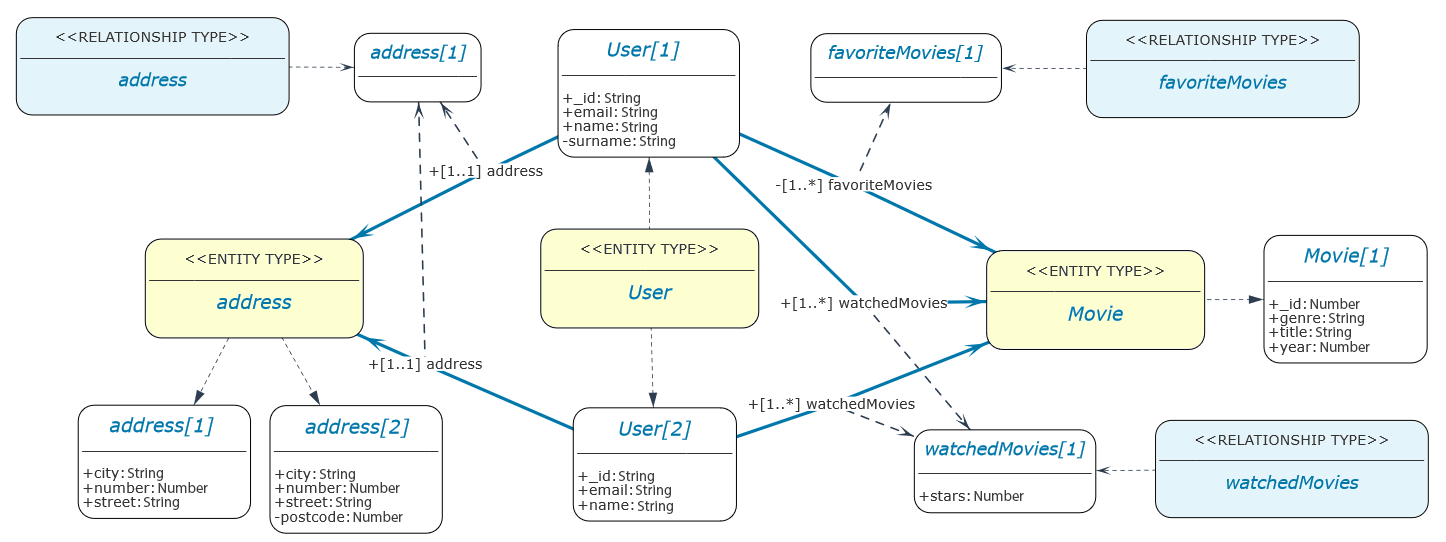}
  \caption{\uschema{} complete schema for ``UP-graph.''}
  \label{fig:q2GraphsAll}
\end{figure*}

Figures~\ref{fig:q1DocAll} and~\ref{fig:q2GraphsAll} show the
\uschema{} complete schemas extracted for ``UP-aggregate'' and
``UP-graph'', respectively.
Both schemas can be obtained with the query ``\texttt{FROM~*~TO~*}'', as
discussed later in Section~\ref{sec:QR}.

In Figure~\ref{fig:q1DocAll}, the schema includes two root entity types:
\texttt{User} and \texttt{Movie}, and two aggregated entity types:
\texttt{Address} and \texttt{WatchedMovies} which are embedded into
\texttt{User}.

\texttt{User} has two variations: \texttt{User[1]} only includes the shared
attributes (\texttt{email}, \texttt{name}, and \texttt{\_id}), while
\texttt{User[2]} has the \texttt{surname} specific attribute and the
\texttt{favoriteMovies} specific reference. Both variations aggregate
\texttt{Address}, but each of them a different \texttt{Address}'s
variation. \texttt{Address} has three shared properties: \texttt{city},
\texttt{number}, and \texttt{street}. Depending on whether the
\texttt{postCode} optional property is present or not, two variations exist
for \texttt{Address}.

In Figure~\ref{fig:q2GraphsAll}, the schema includes three relationship
types: \texttt{address}, \texttt{watchedMovies}, and
\texttt{favoriteMovies}, and two entity types: the \texttt{User} and
\texttt{Movie}. Relationship types \texttt{address} and
\texttt{watchedMovies} result from the aggregated entity types with the
same name in the previous schema. As observed, each reference arrow is
connected to the relationship type that features it, e.g., the two existing
\texttt{watchedMovies} references are connected to the only variation of
the \texttt{watchedMovies} relationship type. Note that this type includes
the attribute \texttt{stars}.

\begin{figure*}[!htb]
  \centering
  \includegraphics[width=0.9\textwidth]{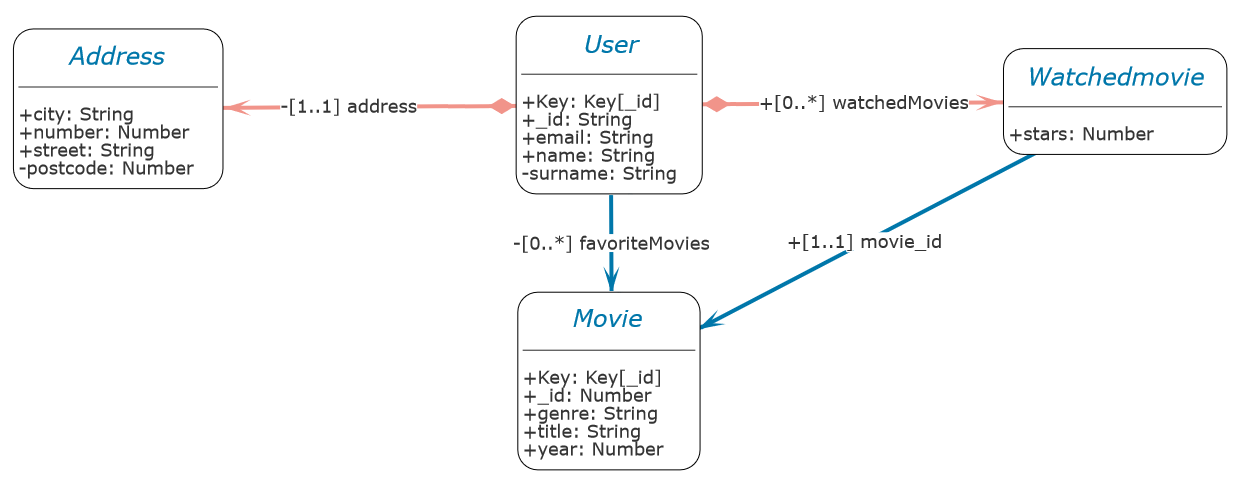}
  \caption{The complete schema of union types for ``UP-aggregate'' schema.}
  \label{fig:unionSchema}
\end{figure*}

Finally, Figure~\ref{fig:unionSchema} shows the complete schema of union
types for ``UP-aggregate'',
which would be obtained with the query \texttt{UNION FROM~* TO~*}.

\section{\skiql{} Query Language: Syntax and Semantics}

\skiql{} was designed to be easy to learn, understand, and write. To
achieve these characteristics, our choice was to create a command language
and visualize the result of the queries in form of a schema graph. Users
should interactively write queries in a console with the results returned
immediately. Being a command language encompasses other advantages, such as
easily extending it with new query commands. \skiql{} is intended for any
stakeholder involved into the development of NoSQL database applications,
such as database administrators, developers, and testers. We considered
that these users could be interested in two kinds of
queries:~(i)~recovering information on properties and variations of a
particular entity type (and relationship type in the case of graph
schemas), and~(ii)~checking the existing relationships (aggregations and
references) among entity types. Both kinds of queries should return a
sub-graph of the database schema. Also, the language should allow each
previously defined schema and sub-schema type to be obtained. It is
important to note that \skiql{} is a DSL aimed to help database
stakeholders to explore large database logical schemas, but it is not
intended to express every possible query on a database schema.

To support the desired queries, \skiql{} provides two kinds of declarative
query statements: ``query on one schema type,'' and ``query on the path
between schema types.'' Next, these query statements are described, and
examples of queries on the \texttt{User Profile} schema will be shown to
illustrate the application and usefulness of \skiql{}.

\subsection{Querying schema types\label{sec:QT}}

\begin{figure*}[!h]
\centering
\begin{minipage}[c]{.45\textwidth}
  \centering \includegraphics[width=0.9\textwidth]{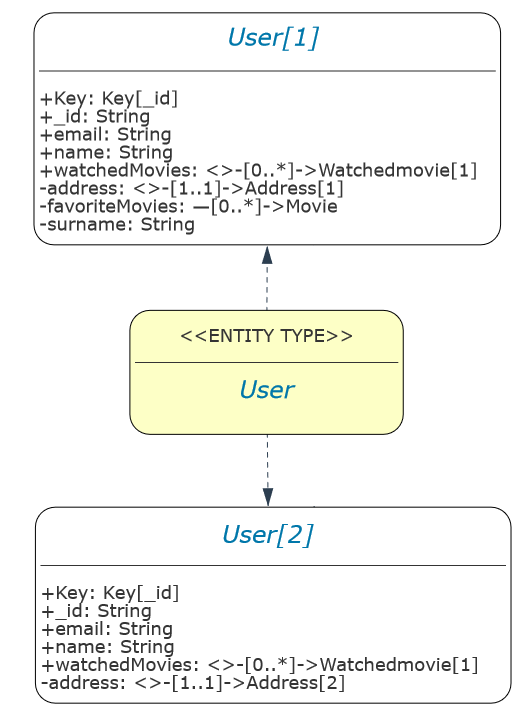}
  \caption{The {\tt User} entity type variation subschema for
    ``UP-aggregate.''}
  \label{fig:entityUser}
\end{minipage}
\qquad
\begin{minipage}[c]{.45\textwidth}
  \vspace{2.2cm}
  \includegraphics[width=0.7\textwidth]{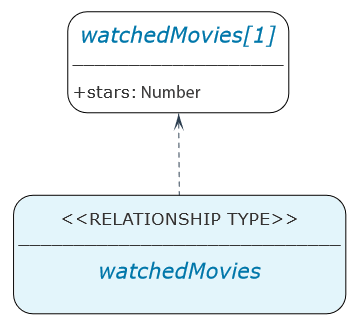}
  \vspace{2.2cm}
  \caption{The {\tt Watchedmovies} relationship type variation subschema
    for ``UP-graph.''}
  \label{fig:relWatchedmovies}
\end{minipage}%
\end{figure*}

A ``query on one schema type'' (QT) allows the user to express a predicate
on a schema type in order to extract information from its \textit{type
  variations sub-schema}. More formally said: Given a schema $S$, a
\textit{schema type query} $qt$ expresses an entity or relationship type
specification $spec$ that conveys a predicate $P(t)$ to be satisfied by the
schema type $t$ of $S$. Such specification consists of a partial
intensional definition of $t$ (a partial list of their features expressed
with its name and optionally its data type),
e.g.,~\texttt{User[name:string, favoriteMovies]}. If such a schema type $t$
exists, the query returns the subgraph of the type variations sub-schema
that satisfies the predicate. In this case, the relationships are enclosed
in the lower compartment of the variations, as indicated above, and can be
observed in Figure~\ref{fig:entityUser} that is commented below.

Regarding the syntax, a QT statement consists of three parts. First, a
keyword indicating the schema type on which the query is applied:
\texttt{ENTITY} for an entity type, \texttt{REL} for a relationship type,
and \texttt{ANY} (both schema types). Next, the name of the schema type,
which can be followed of an optional \emph{variation filter} clause. The
type name can be expressed in different ways: the exact name to be matched,
\texttt{*} to get all entities, use the symbol ``{\tt *}'' to establish a
prefix, suffix or both, e.g., \texttt{*Movie} to express an entity type
name ending in ``Movie'',
or Java-like regex expressions.

An excerpt of the EBNF grammar of this kind of query is the following:

\begin{grammar}
<type-query> ::= [`UNION'] (`ENTITY' | `REL' | `ANY') <TypeSpec>
                 [<variation-filter>] | [<operations>]

<TypeSpec> ::= [`*']<typeName>[`*']|`*'|`r"' <regexp> `"'

<variation-filter> ::= `[' <feature> {`,' <feature>} `]'

<feature> ::= <nameFeature> `:' [<featureType>]

<featureType> ::= <AttributeDataType> | <AggregatedType> | <ReferenceType> | `?'

<AttributeDataType> ::= <BasicType> | <CollectionType>

<BasicType> ::= `number' | `string' | `boolean'

<CollectionType> ::= <BasicType> `[' `]'

<AggregatedType> ::= `AGGR` `<' <typeName> `>'

<ReferenceType>  ::= `REF` `<' <typeName> `>'

<operations> ::= <operation> {`,' <operation>}

<operation> ::=  `keys' | <date-interval>

<date-interval> ::=   `history' ('before' <date> | `after' <date> | `between' `('<date> `,' <date>`)')
\end{grammar}

An \emph{entity type variations subschema} or a \emph{relationship type
  variation subschema} is returned when the \texttt{ENTITY} or \texttt{REL}
keywords are followed by the name of an entity or relationship type,
respectively. Figure~\ref{fig:entityUser} shows the results obtained for
the query ``\texttt{ENTITY User}'' and Figure~\ref{fig:relWatchedmovies}
for ``\texttt{REL watchedMovies}'' query. Note that two relationship type
variation subschemas would be obtained for the query ``\texttt{REL
  Movie}'', those that corresponds to the \texttt{watchedMovies} and
\texttt{favoriteMovies} relationship types. \texttt{ANY} keyword is used to
express that the name can refer to either an entity type and a relationship
type, e.g. this would occur if ``\texttt{ANY Address}'' is issued on the
\emph{User Profile} graph database. These three forms of query could be
used to check if a schema type is present or not in the schema.

A \textit{variation filter} enumerates the list of features that a
variation must have in order to be selected, i.e. a QT query predicate.
Each feature is specified by indicating its name and type separated by a
colon, and features are separated by commas. The data types allowed for
attribute features are \texttt{Number}, \texttt{String}, \texttt{Boolean},
as well as \emph{collections} of values of these data types. The
collections are those included in the \uschema{} metamodel: Arrays, Sets,
Lists, Tuples, and Maps. An array contains values of the same type, and the
array type is specified by adding square bracket after the type name, for
example \texttt{String[]}. The rest of collections are specified with the
collection type name followed by the base type between angle brackets, for
example, \texttt{Set<String>} and \texttt{Map<String, Number>}. A question
mark can be used to indicate that the property type is unknown or either
can be omitted. The types for relationship features are expressed with the
prefix ``AGGR'' for aggregates or ``REF'' for references.

Query \emph{Q1} shows a variation filter example: ``find all \texttt{Users}
variations with the \texttt{name:~String} attribute, and another feature
named \texttt{favoriteMovies} whose type is unknown.'' The result returned
would be the same as Figure~\ref{fig:entityUser} but not including the
variation \texttt{User[2]} that does not meet the filter.

\begin{lstlisting}[language=skiql,frame=tb,title={Q1},label=Q1]
ENTITY User [name: string, favoriteMovies]
\end{lstlisting}

A QT statement can also include operations that are applicable on schema
types or variations. These operations follow the type name or filter. At
this moment, two operations have been defined: ``\texttt{keys}'' returns
the keys of the specified entity types, and ``\texttt{history}'' returns a
graph that shows the timeline of appearance of variations in a given date
interval.

In a variation filter, the \texttt{shared}, \texttt{non-shared}, and
\texttt{specific} keywords can be used when specifying a feature. For
example the query ``\texttt{ENTITY * [shared id]}'' would return all the
entity types having a shared property named ``\texttt{id}'' of unknown data
type, and ``\texttt{ENTITY User [shared surname: string]}'' would return all
\texttt{User} variations having a shared property named \texttt{surname} of
data type {\tt String}. In the previous section, we showed the use of QT
queries to obtain complete entity and relationship schemas:
``\texttt{ENTITY~*}'' and ``\texttt{REL~*}''.

\subsection{Querying aggregations and references\label{sec:QR}}

As explained in Section~\ref{sec:uschema}, two kinds of relationships can
be found in NoSQL stores: \textit{aggregations} from a origin variation to
another target variation (only in aggregate-based systems), and
\textit{references} from a origin variation to a target entity type. A
\emph{relationship query} (QR) selects the sub-schema that includes the
specified relationships in the query. This kind of query can be formally
defined as follows. Given a schema $S$, a \textit{relationship query} $qr$
expresses a specification $oe$ of a entity type $t$ belonging to $S$, and
one or more relationship specifications $r_i, i=1 \ldots n$, each of them
indicating a relationship kind $k_i$ and a specification of a target entity
type $tt_i$. Thus, $qr$ formulates a predicate $P(t,r)$ that is formed by a
conjunction of logical operands, and each operand expresses that the
relationship of kind $k_i$ exists from $t$ to $tt_i$. If this predicate is
satisfied, the query returns the subgraph of $S$ that contains the set of
relationships $r_i$. While a QT query retrieves a subgraph of a type
variations sub-schema, a QR query may return any subgraph of a complete
schema.

Regarding the syntax, a QR query consists of a \texttt{FROM} clause
followed by a \texttt{TO} clause. The former specifies the source type of
the relationship, and the latter the target type and the kind of
relationship. Depending on whether the prefix \texttt{UNION} is present or
not, variations or union schema types are returned as source and target of
the relationship returned.

The syntax is expressed below in form of an EBNF grammar.

\begin{grammar}
<schema-query> ::= [`UNION'] <from-clause> <to-clause>

<from-clause> ::= `FROM' (<entitySpec> [<variation-filter>] | `_')

<to-clause> ::= `TO' <rel-spec> {`,' <rel-spec>}

<rel-spec> ::= [`>>'] <entitySpec> [<variation-filter>]
                         (`REF' [<featureName>] [<variation-filter>] |\\
                          `AGGR' [<featureName>] | `ANY' [<featureName>]) | `_'

<entitySpec> ::= [`*']<typeName>[`*']|`*'|<regexp>)
\end{grammar}

A \emph{from clause} is formed by the \texttt{FROM} keyword followed by an
entity type name, and an optional variation filter that is expressed with
the syntax exposed for QT queries. An empty \texttt{FROM} clause
(underscore symbol) denotes that no entity type or variation could have a
relationship to the target entity type.

A \emph{to clause} is formed by a list of relationship specifications which
are pairs formed by an entity type and a keyword denoting the kind of
relationship: ``\texttt{REF}'', ``\texttt{AGGR}'', or ``\texttt{ANY}''.
This latter keyword is used to indicate that the relationship can be
aggregation or reference. A variation filter can only be used with
\texttt{AGGR}. A star can be used to refer to ``any entity type'' and an
underscore to ``no entity type.''

\begin{figure*}[!htb]
  \centering
  \includegraphics[width=0.8\textwidth]{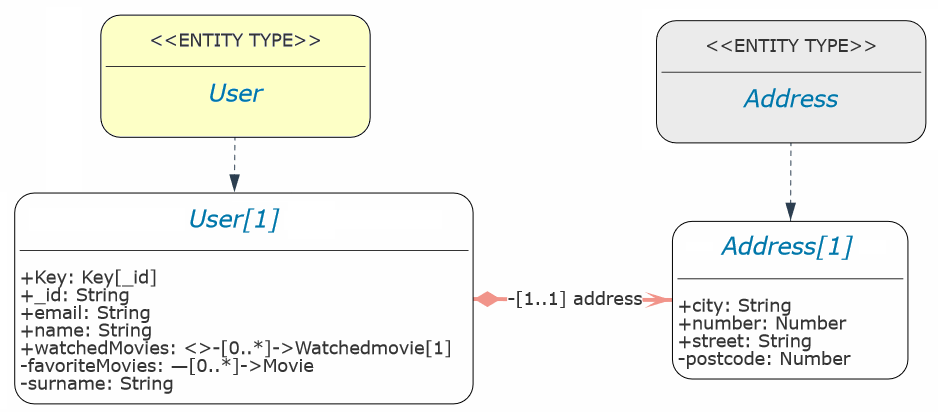}
  \caption{Subschema returned for queries Q2 and Q5 on ``UP-aggregate''
    schema.}
  \label{fig:q4UserToAddress}
\end{figure*}

\emph{Q2} is a QR query specifying the condition ``aggregation between
\texttt{User} variations including the attribute {\tt surname} of type
String and an \texttt{Address} variation.''
As shown in Figure~\ref{fig:q4UserToAddress}, if the query is applied on
the ``UP-aggregate'' database schema,
the result is the \texttt{User} variation ``{\tt User[1]}'', which appears
connected to the \texttt{Address} variation ``{\tt User[1]}'' through an
aggregation relationship. Note that the \texttt{watchedMovies} aggregation
or \texttt{favoriteMovies} reference are not shown in form of edge as they
are not part of the set of relationships satisfying the query predicate, so
they appear in the variation box as features.

\begin{lstlisting}[language=skiql,frame=tb,
%title={Q2. Returns those \texttt{User} variations
%aggregating \texttt{Address} objects and having an attribute named ``surname.''}
title={Q2},
%Check if any
%\texttt{User}'s variation containing a \texttt{surname:string}
%property and is connected to the \texttt{Address} entity by means of a
%aggregation.},
label=Q2]
FROM User[surname:string]
TO Address AGGR
\end{lstlisting}

\begin{figure*}[!htb]
  \centering
  \includegraphics[width=\textwidth]{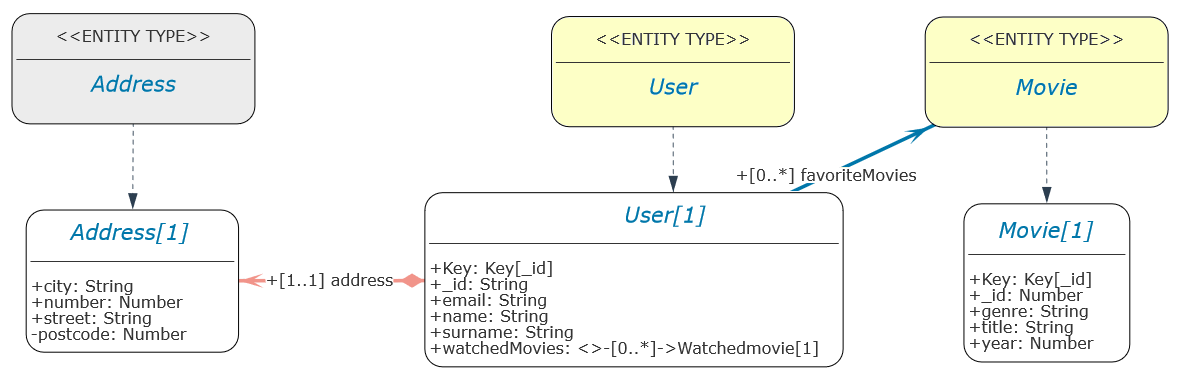}
  \caption{Subschema returned for query Q4 on ``UP-aggregate'' schema.}
  \label{fig:q8UserRefAggr}
\end{figure*}

Table~\ref{tab:schemaLevelQueryExamples} shows more QR query examples for
the \texttt{User Profile} schema. \emph{Q3} checks if the \texttt{User} entity
type has incoming relationships, and would return a message indicating that
\texttt{User} is not target type of any relationship. \emph{Q4} checks if
\texttt{User} has references to \texttt{Movie} and aggregations to
\texttt{Address}, and would return the \texttt{User[1]} variation connected
to \texttt{Movie} and \texttt{Address} through the \texttt{favoriteMovies}
reference and the \texttt{address} aggregation, respectively, as shown in
Figure~\ref{fig:q8UserRefAggr}. Note that \texttt{watchedMovies}
aggregation is not shown in form of an edge for the reason explained above.

\begin{table*}[h!]
\centering
\begin{tabular} {ll}
  \begin{minipage}[t]{.45\textwidth}%
\begin{lstlisting}[language=skiql,label=Q3,title=Q3,frame=tb]
FROM _
TO User
\end{lstlisting}
\end{minipage}
            &
              \begin{minipage}[t]{.45\textwidth}%
\begin{lstlisting}[language=skiql,label=Q4,title=Q4,frame=tb]
FROM User
TO Movie REF, Address AGGR
\end{lstlisting}
              \end{minipage}\\
  \begin{minipage}[t]{.45\textwidth}%
\begin{lstlisting}[language=skiql,label=Q5,title=Q5,frame=tb]
FROM User [favoriteMovies]
TO Address [postcode] AGGR
\end{lstlisting}
  \end{minipage}
            &
              \begin{minipage}[t]{.45\textwidth}%
\begin{lstlisting}[language=skiql,label=Q6,title=Q6,frame=tb]
FROM User
TO >> Movie
\end{lstlisting}
              \end{minipage}
\end{tabular}
\caption{QR query examples.}
\label{tab:schemaLevelQueryExamples}
\end{table*}

\emph{Q5} query retrieves relationships whose origin are \texttt{User}
variations having \texttt{favoriteMovies} feature of unknown type, and the
target is an \texttt{Address} variation containing the attribute
\texttt{postcode} through aggregation. The returned diagram is the same as
for query \emph{Q2}, which is shown in Figure~\ref{fig:q4UserToAddress}.

\begin{figure*}[!htb]
  \centering
  \includegraphics[width=\textwidth]{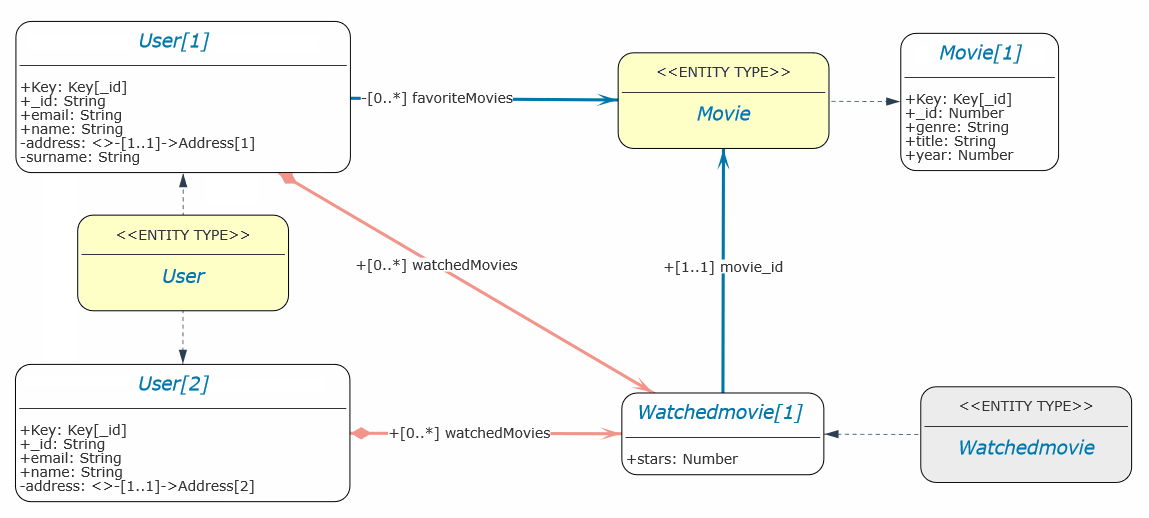}
  \caption{Subschema returned for query \emph{Q6} on ``UP-aggregate''
    schema.}
  \label{fig:q9Indirect}
\end{figure*}

In the query execution, relationships specified
are direct by default, but a path of any length can be indicated by using
the \texttt{>>} prefix, as illustrated in query \emph{Q6}. This query
checks if the \texttt{User} entity is connected to \texttt{Movie} by means
of a path that can include any number of aggregations and references.
Figure~\ref{fig:q9Indirect} shows the subschema returned where the
\texttt{User[1]} variation is directly connected to \texttt{Movie} ({\tt
  favoriteMovies} reference), and \texttt{User[2]} variation is indirectly
connected to\texttt{Movie} through the \texttt{WatchedMovies} aggregate
entity type that references \texttt{Movie}.

All the relationships directly or indirectly incoming/outgoing to/from a
given entity type can be obtained by using ``{\tt *}'' to specify the
schema type name, as shown below for the \texttt{User} entity type.

\begin{itemize}
\item \texttt{FROM~User~TO~*} returns all relationships outgoing from
  \texttt{User},
\item \texttt{FROM~*~TO~User} returns all relationships incoming to
  \texttt{User}.
\item \texttt{FROM~User~TO~>>~*} returns all relationships outgoing from
  \texttt{User} to any entity type connected directly or indirectly.
\item \texttt{FROM~*~>>~TO~User} returns all direct or indirect
  relationships incoming to \texttt{User}.
\end{itemize}

These four queries return union types instead variations if the
\texttt{UNION} prefix is present.

\begin{figure*}[!htb]
  \centering
  \includegraphics[width=0.8\textwidth]{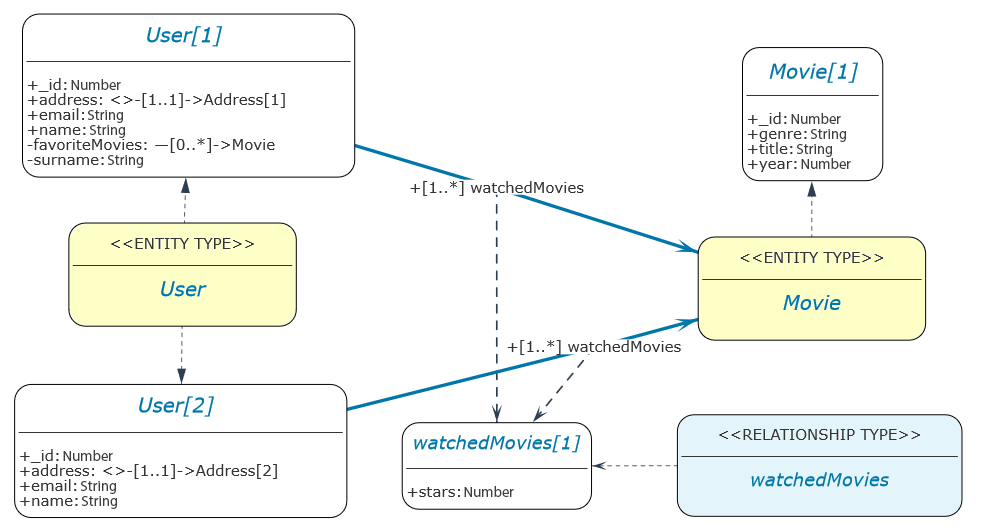}
  \caption{Subschema returned for query \emph{Q7} on ``UP-graph'' schema.}
  \label{fig:q10RefWithAttributes}
\end{figure*}

In the \texttt{TO} clause, the keyword indicating the kind of relationship
can be optionally followed by the name of a property, so that the query
would only return relationships with that name. As references can have
attributes in graph systems, they are instances of relationship types,
\skiql{} allows variation filters to be used to specify relationship
attributes. Query \emph{Q7} would check if \texttt{User} is connected to
\texttt{Movie} through a reference which has a \texttt{stars} attribute of
type \texttt{Number}. The result of this query issued on ``UP-graph''
schema is shown in Figure~\ref{fig:q10RefWithAttributes}.

\begin{lstlisting}[language=skiql,frame=tb,
title={Q7},
label=Q7]
FROM User
TO Movie REF [stars: Number]
\end{lstlisting}

\begin{figure}[!htb]
  \includegraphics[width=\textwidth]{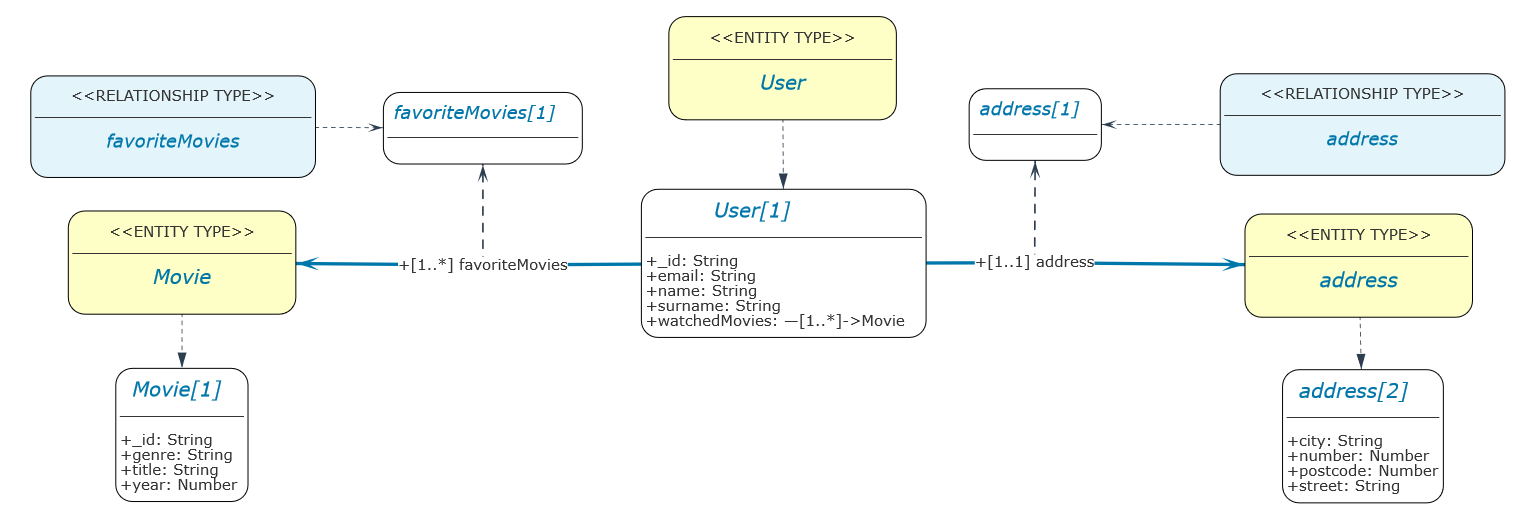}
  \caption{Subschema returned for query \emph{Q8} on ``UP-graph'' schema.}
  \label{fig:q8UserToMovieToAdressVar}
\end{figure}

Finally, \emph{Q8} shows a QR query issued on ``UP-graph'' to find if the
schema contains \texttt{User} entity type variations with the
\texttt{surname} attribute, which are connected both to \texttt{Address}
variations with \texttt{postcode} and to \texttt{Movie} through only
\texttt{favoriteMovies} references.
Figure~\ref{fig:q8UserToMovieToAdressVar} shows the result obtained for
\emph{Q8}.

\begin{lstlisting}[language=skiql,frame=tb,title={Q8},label=Q8]
FROM User [surname: string]
TO Address [postcode],
   Movie REF favoriteMovies
\end{lstlisting}

In addition to the graphical notation, the results can be also displayed as
a set of tables, one for each returned schema type. Each table has a row
for each variation, and rows have four columns: schema type name, variation
identifier, number of instances, and a listing of features expressed in the
format used in the diagrams. This textual notation can be useful if the
number of variations returned is high.

\section{Implementation of \skiql{}\label{implementation}}

\skiql{} was created with a metamodel-based language workbench,
Xtext~\cite{bettini2016}. As it is well-known~\cite{volter-book}, these
tools automate the building of DSLs by automatically generating an editor,
a parser, and a model injector from the EBNF-like grammar or metamodel of
the language.

Once the syntax of \skiql{} was determined, a metamodel-based
approach~\cite{greenfield-2004} was applied to implement the language. We
first defined the metamodel (i.e., its abstract syntax)
and then wrote the grammar in form of Xtext syntax rules. A translational
approach~\cite{greenfield-2004} was applied to define the \skiql{}
semantics: \skiql{} queries are written with the generated editor, and the
model injector automatically produces \skiql{} models in Ecore/EMF
format~\cite{steinberg-emf2009}. Then, a query interpreter has as input
these query models along with the \uschema{} model that represents the
NoSQL schema on which queries are issued.

Every time a QT or QR query is issued, the interpreter launches the
injector execution to convert the query script into a \skiql{} model. Then,
the interpreter performs the following two-step process. Firstly, the query
model is analyzed to identify the conditions to be satisfied, and then the
\uschema{} model (i.e. the schema) is traversed to obtain the elements to
be returned in the result graph. This second step is a model-to-model
transformation that extract the part of the \uschema{} model that
constitute the desired result. Note that the \uschema{} model is both the
source and target metamodel in this transformation. The interpreter has
been implemented with the language Xtend~\cite{bettini2016}.

\begin{figure*}[!htb]
  \centering
  \includegraphics[width=\textwidth]{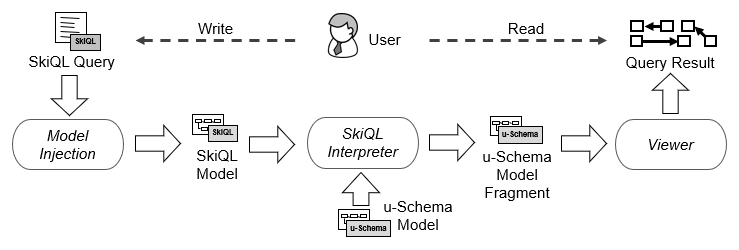}
  \caption{An overview of the visualization process and its
    implementation.}
  \label{fig:process}
\end{figure*}

As shown in Figure~\ref{fig:process}, the interpreter has been integrated
with a \skiql{} schema viewer to graphically show the result of the
\skiql{} queries that are written in the console (i.e.,~the generated
editor). This viewer receives as input the \uschema{} model produced by the
interpreter, and then creates the corresponding graph by applying the
mapping exposed in Table~\ref{table:uschema-viewer-mappings} between
\uschema{} elements and graphical notation elements.

The viewer has been implemented by using VisJS\footnote{VisJS
 Webpage:~\url{http://visjs.org}.} as graphical visualization API. Our
tool (query interpreter and schema viewer) is a Web application that allows
queries to be entered through an editor and also visualized in the browser.

\section{Evaluation}
\label{sec:evaluation}

In this section, we will present the evaluation of \skiql{}, which has been
carried out in two forms. First, we measured some language metrics for
\skiql{} and compared the results with those of other query languages.
Secondly, we surveyed some experienced NoSQL researchers on \skiql{}
features.

\subsection{Calculating Language Metrics}
\label{metrics}

In order to assess to what extent \skiql{} is a simple and easy to learn
language, we calculated the metrics defined
in~\cite{vcrepinvsek2010automata}, and compared the results obtained to
those of other three query languages, in particular: Cypher, SPARQL, and
GraphQL. Table~\ref{tab:metrics} shows the results for these four
languages.\footnote{The {\tt cfgMetrics} program was executed to calculate
  language metrics.}

\begin{table*}[]
\begin{center}
\begin{tabular}{llrrrr}
\toprule
\textbf{Metric}& \textbf{Definition} & \textbf{\skiql{}} & \textbf{GraphQL} & \textbf{Cypher} & \textbf{SPARQL} \\
\midrule
TERM   & Terminals & 29        & 44          & 115        & 158    \\

VAR    & Non-Terminals & 39        & 71          & 99          & 123    \\

HAL    & Designer effort for grammar understanding & 12.81   & 43.03     & 141.16   & 116.55 \\

LRS    & Complexity independent size & 310      & 603        & 15756    & 15172  \\

LAT/LRS & Ease of understanding & 0.085   & 0.155     & 0.138     & 0.160 \\
\bottomrule
\end{tabular}
\end{center}
\caption{\skiql{}, GraphQL, Cypher and SPARQL metrics.\label{tab:metrics}}
\end{table*}

The metrics used measure the following quantities. \textit{TERM} and
\textit{VAR} the number of terminals and non-terminals, respectively.
\textit{HAL (Halstead metric)} the designer effort to understand
  the grammar. \textit{LRS} the complexity of the language independent
of its size, and \textit{LAT/LRS} measures the ease of understanding the
language.

Since queries were translated to Cypher in a first version of \skiql{}, we
have considered this graph query language. SPARQL and GraphQL were chosen
as they are widely used query languages. The former is the standard RDF
query language, and GraphQL is increasingly used to query APIs in web and
mobile applications. The structure of GraphQL types is similar to
\uschema{} entity types but structural variations are not allowed. SQL was
not considered because it is a large language whose specification has a
large number of statements that are not used to query data.

Analyzing the results obtained for each language, we found that Cypher and
SPARQL are larger languages than GraphQL and \skiql{}, as they are intended
to query more complex data structures. This is shown by the TERM and VAR
values in Table~\ref{tab:metrics}. The TERM values are very close for
\skiql{}~(29) and GraphQL~(44), while Cypher~(115) and SPARQL~(158) have
higher values. With non-terminals, the VAR value for \skiql{} is
significantly lower than the other three values. The metric HAL shows that
Cypher is the more complex~(141.2), followed by SPARQL~(116.5), while
GraphQL~(43.0) and \skiql{}~(12.8) are much simpler languages. \skiql{} is
appreciably the least complex of the four. It is convenient to remark that
Cypher is about ten times more complex than \skiql{}. LRS is other metric
that measures the language complexity, and its values are consistent with
those obtained for HAL. Regarding LAT/LRS, Table~\ref{tab:metrics} shows
that GraphQL~(0.160) and SPARQL~(0.155) are somewhat more difficult to
learn than Cypher~(0.138), and the \skiql{} value is half of the Cypher
one. This difference in the easiness to learn a language is similar to
those calculated for other DSLs defined as alternative to general purpose
languages, such as~\cite{hoyos-dsl2013}. \skiql{} can be considered as a
more abstract language defined on top of Cypher to query database schemas.
With \skiql{}, developers save time writing queries with a simpler and both
easier to understand and to learn language than Cypher to query schema
graphs. As indicated in Section~\ref{sec:relatedwork}, the average number
of LoC for simple queries expressed in Cypher is $8$,
Listings~\ref{lst:cypherQ1} and~\ref{lst:cypherQ4} show Cypher queries for
\emph{Q1} and \emph{Q4} \skiql{} queries.  Queries \emph{Q2} and \emph{Q8}
are challenging to write in Cypher due to the difficulty
of implementing variation filters.
However these two queries are very easy to
write in \skiql{}. As shown in Listings~\ref{lst:cypherQ1}
and~\ref{lst:cypherQ4}, Cypher queries are longer and more
complex than \skiql{} queries. In addition, the user should learn
Cypher and know how schemas are represented as graphs in the database.

\begin{figure}
\begin{lstlisting}[ frame=tb, caption={Cypher query for Q1: ENTITY User {[}
name:string, favoriteMovies {]}.}, label={lst:cypherQ1}, numbers=left,
language=cypher ]
WITH ["name:string", "favoriteMovies"] AS properties
MATCH (p:Property) WHERE p.nameType IN properties
WITH collect(p) AS ps, properties
MATCH (ev:EntityVariation {entity:"Users"})-->(:Property)
WHERE ALL (p IN ps WHERE (ev -->(p))
RETURN ev
\end{lstlisting}
\end{figure}

\begin{figure}
\begin{lstlisting}[ frame=tb, caption={Chyper query for Q4: FROM User TO
Movie REF, Address AGGR.}, label={lst:cypherQ4}, numbers=left,
language=cypher ]
MATCH c = allShortestPaths(
(:ENTITY {name:"Users"})-[:ENTITY_VARIATION|PROPERTY|REFS_TO*1..3]->
(:ENTITY {name:"Movies"}) )
MATCH c2 = allShortestPaths(
(:ENTITY {name:"Users"})-[:ENTITY_VARIATION|PROPERTY|AGGREGATES*1..3]->
(:ENTITY_VARIATION)<--(:ENTITY {name:"Address"}) )
RETURN c, c2
\end{lstlisting}
\end{figure}

\subsection{Survey on \skiql{} Features\label{sec:metrics}}

We surveyed a total number of~31 participants, which had no knowledge on
\skiql{}:~7~researchers from other research groups,~8~Spanish developers
experienced in MongoDB,~6 members of our research group, and~10~students of
a Big Data master.

We provided to the participants a document with three parts: a \skiql{}
tutorial with examples of queries and the result graph, several query
exercises to be solved by respondents, and a questionnaire of six items to
evaluate \skiql{}. The questionnaire is shown in
Table~\ref{tab:table-questionnaire}. Each question had to be assessed with
a mark from~1 to~5 in the \emph{Likert} scale.

The participants were provided with a virtual machine with all the
necessary tools to write and execute SKiQL queries. Therefore, they all
used the same environment and with the same assistants from the editor.
They completed the questionnaire once they solved the exercises. The six
questions asked were about the following features: legibility of result
graphs, ease to learn, ease to understand queries, adequate expressiveness,
usability of the environment, and usefulness of the language.

\paragraph{Results and Discussion}

Table~\ref{tab:table-questionnaire} shows the average and the standard
deviation of the participant's scores for each of the six questions.

\begin{table}[]
\begin{center}
\begin{tabular}{lrr}
\toprule
\textbf{Question} & \textbf{AVG} & \textbf{SD} \\
\midrule
Is \skiql{} easy to learn? &
4.19 & 0.38\\

Is \skiql{} easy to read? &
4.84 & 0.35\\

Is \skiql{} easy to write? &
4.23 & 0.68\\

Is \skiql{} expressiveness appropriate? &
4.26 & 0.79\\

Could \skiql{} be useful to developers? &
3.81 & 0.72\\

Is the visualization understandable? &
4.36 & 0.81\\
\bottomrule
\end{tabular}
\end{center}
\caption{Questionnaire results.\label{tab:table-questionnaire}}
\end{table}

\textbf{Easy to learn} None of the respondents considered the language
difficult to learn,
the average obtained for this question is~4.19 with a standard deviation
of~0.38, supporting \skiql{} is easy to learn. In their comments, the
participants indicated that the documentation provided had been very
useful.

\textbf{Usability} This feature includes the ease of reading and
understanding queries (second question of the survey) as well as the ease of
writing queries (third question). The former is the best evaluated in the
survey,
with an average of~4.84, and the participants strongly agreed that \skiql{}
is a simple and concise language. On the other hand, writing is well
evaluated~(4.23) but not so well as reading~(4.84), although none of the
participants scored negatively. Their scores divided equally between the
two positive positions.

\textbf{Expressiveness} The expressiveness of the language refers to
whether the language offers all the needed features. This characteristic is
usually considered with the preciseness feature to measure the
effectiveness: its ability to perform complex queries with the least number
of elements. In our case, we only considered expressiveness because
\skiql{} is a command language, and a single query is executed each time.
Most of the participants positively positioned on expressiveness with an
average of~4.26.

\textbf{Usefulness of language} In the usefulness of the language, their
scores divided equally between the neutral and positive answers (average
is~3.81). More than half of respondents claimed to agree or strongly agree,
and the other remaining took a neutral position. Some of them pointed out
that the language had a great similarity in simplicity to the SQL language.

\textbf{Legibility of the result graph} The legibility of the graph
returned refers to aspects such as the proper understanding of the schema
in form of a graph, the ability to know the kind of each shown property as
well as the variation which it belongs to, and understanding the
relationships between different entities, among others. None of the
participants considered that the visualization representation is illegible,
and their scores divided equally between the neutral and positive
positions. Thus most
respondents are positively positioned regarding to the understanding of
the graph that represent result schemas with an average of~4.36.

\textbf{Limitations of the validation} Among the limitations of the
validation is that large schemas were not used, in an attempt to reduce the
effort to learn \skiql{}. The survey included a limited number of
exercises. We included very simple exercises to familiarize respondents
with \skiql{}, and then we wanted to assess the expressiveness of the
language, and also to cover most of the features of the language, so we
included more complex queries. This may have lead some participants to
consider using the language to be slightly difficult. In the documentation
given to respondents, we included a brief explanation on the metamodel used
to represent schemas, and some of them had problems to solve exercises
because they had not clear notions of structural variation of an schema
type.

\section{Conclusions and Further Work}\label{sec:conclusions}

In this paper, we presented the \skiql{} generic language capable of
querying logical schemas of NoSQL and relational databases. In addition to
the common elements of logical schemas (entities, properties, aggregation,
references, and keys), the notion of structural variation has also been
taken into account in its design. When extracting NoSQL schemas, variations
of entity types can be represented as union types, or either the proper
variations can be added to the discovered schema. Our query language allows
for both possibilities. On the other hand, when designing schemas,
variations could be useful, for instance, if a hierarchy of entity types is
part of the conceptual schema~\cite{alberto-subtypes2021}. We have also
considered relationship types for graph schemas.

Although the focus of our work is schema querying, a graphical
visualization has also been devised for query results. The \uschema{}
unified metamodel has been used to represent logical schemas, but the
\skiql{} language and the visualization are independent of the kind of
schema representation.

\skiql{} is easy to learn and usable as showed the evaluation performed,
and it allows schemas to be easily explored. A possible strategy for
exploring an unknown schema might be to first obtain the complete schema of
union types. Then, each schema type could be consulted to know its
variations. Once that knowledge is acquired, queries on relationships could
be issued to find more specific information.

It is remarkable that as far as we know, \skiql{} is the firs language
proposed to query NoSQL logical schemas, an utility widely available for
relational schemas.

Finally, we are planning to extend \skiql{} to allow queries on elements
that are specific to physical schemas, such as indexes, deployment, and
sharding information. The statements to be added to the language should be
based on an extension of \uschema{}. Instead of adding elements to
\uschema{}, the definition of a separated generic physical metamodel could
be convenient, and both metamodels should be linked.

\bibliographystyle{plainurl}
\bibliography{ms}

\end{document}